\documentclass[preprint,proceedings]{rmaa}

\usepackage{paralist}

\usepackage{psfrag,color}

\usepackage{epsf}
\usepackage{psfig}
\usepackage{graphicx}
\newcommand{\LZ}{\mathcal{L}}

\bibpunct[]{(}{)}{;}{a}{}{,} 

\SetYear{2006}
\SetConfTitle{XIth IAU Regional Meeting (Invited Review)}

\title{Near-Field Cosmology with Horizontal Branch\\ and RR Lyrae Stars} 

\author{
  M. Catelan\altaffilmark{1} 
  }

\altaffiltext{1}{Pontificia Universidad Cat\'olica de Chile, Santiago, Chile.}

\shortauthor{Catelan}
\shorttitle{Near-Field Cosmology with HB and RR Lyrae Stars}

\fulladdresses{
\item Pontificia Universidad Cat\'olica de Chile, 
  Departamento de Astronom\'\i a y Astrof\'\i sica, Av. Vicu\~na 
  Mackenna 4860, 782-0436 Santiago, Chile
  (\email{mcatelan@astro.puc.cl}).
}

\listofauthors{M. Catelan}
\indexauthor{Catelan, M.}

\abstract{The importance of horizontal branch and RR Lyrae stars is 
  discussed in the context of cosmological arguments for the formation 
  of the Galactic halo and its satellite dwarf galaxies. It is shown, in particular, 
  that the Galactic halo globular cluster system cannot have formed from the 
  accretion of ``protogalactic fragments'' resembling the very early counterparts 
  of the present-day dwarf satellite galaxies of the Milky Way, or else its 
  RR Lyrae properties would be very different from what is currently observed.}

\resumen{Se discute la importancia de estrellas de rama horizontal y variables
  RR Lyrae en el contexto cosmol\'ogico de la formaci\'on del halo Gal\'actico
  y de sus galaxias-sat\'elite. Se muestra, en particular, que el sistema
  Gal\'actico de c\'umulos globulares no puede haberse formado a partir de la 
  incorporaci\'on de ``fragmentos protogal\'acticos'' similares a las 
  contrapartes primordiales de las galaxias enanas sat\'elites de la 
  V\'\i a L\'actea, porque, de ser as\'\i , las propiedades de las variables
  RR Lyrae en aqu\'el sistema resultar\'\i an muy distintas a lo que hoy se 
  observa.}

\addkeyword{Galaxies: Dwarf}
\addkeyword{Galaxy: Formation}
\addkeyword{Galaxy: Globular Clusters: General}
\addkeyword{Stars: Horizontal-Branch}
\addkeyword{Stars: Variables: Other}

\begin{document}
\maketitle

\section{Introduction}
\label{sec:int}

How did the Galactic halo form? Modern 
$\Lambda$CDM cosmology favors a hiearchical picture much like 
the one envisaged by \citet{sz78}, with a galaxy like the Milky Way being 
the process of merger and accretion of hundreds of smaller entities
\citep[e.g.,][]{maea03} not unlike the dwarf satellite galaxies that 
are still seen orbiting the 
Galaxy today. Indeed, there is at least one well-documented 
example of a dwarf galaxy---the Sagittarius dwarf spheroidal 
(dSph)---being currently accreted by the Milky Way \citep{iea95}. 
On the other hand, a significant body of evidence points, perhaps rather 
surprisingly, to a scenario which appears largely inconsistent with 
$\Lambda$CDM predictions.

\begin{figure}[ht]
  \includegraphics[angle=0,scale=0.475]{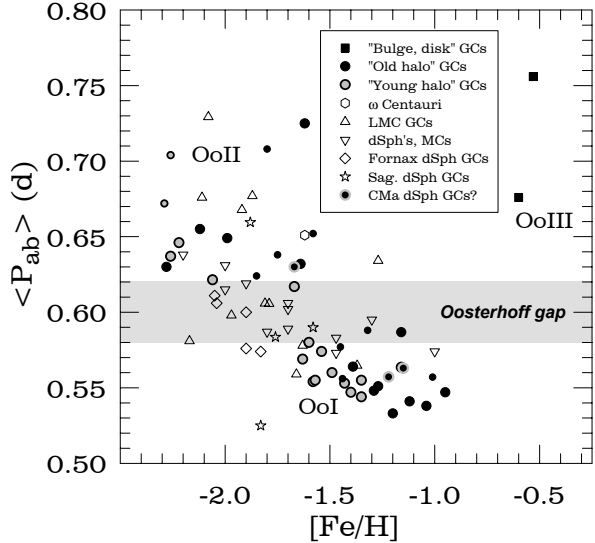} 
  \caption{
  Systematics of the Oosterhoff dichotomy. Note that it is clearly 
  present among bona-fide Galactic globular clusters, but not 
  among the Milky Way dwarf satellite galaxies and their globular 
  clusters. 
\label{fig:OO}
}
\end{figure}

Indeed, much of the current evidence appears to suggest that dwarf galaxies 
such as the ones currently orbiting the Milky Way cannot have been primarily 
responsible for the formation of the Galactic halo
\citep[see also][ for evidence favoring monolithic collapse in the case of 
Coma cluster galaxies]{dfea05}. Among the better known inconsistencies between 
the modern hierarchical paradigm and the 
empirical evidence are the following: i)~The Galactic halo contains 
but a few stars younger than the bulk of the halo population, unlike 
most of the Milky Way's satellite dSph galaxies which often do contain sizeable 
young components---thus suggesting that dSph galaxies cannot have been the 
primary ``building blocks'' of the Milky Way \citep*{muea96}. 
ii)~The detailed abundance patterns among stars in dwarf satellite galaxies 
\citep*[][]{msea03,etea03,kvea04,dgea05,bpea05}
is strikingly different from that in the Galactic halo, again suggesting that 
the latter cannot have been built up from protogalactic fragments resembling 
the former. 

However, most such objections to the hierarchical model for the formation of 
the Milky Way can be avoided if the vast majority of the 
accretion events took place {\em very early 
on} in the Galaxy's history \citep[e.g.,][]{afea06,eg06}. In this scenario, 
the satellites that survived to this day have undergone additional chemical 
enrichment {\em over a prolonged timespan}. 
For these reasons, in order to place meaningful constraints on the way our 
(undoubtedly old) Galactic halo formed, we should really compare the 
{\em very oldest stars} in both the present-day halo and the Milky Way 
dwarf satellite galaxies. 

RR Lyrae stars, as unmistakable tracers of the oldest populations 
of galaxies, provide us with an excellent means to probe into these earliest 
stages of the Galaxy's formation history. In particular, if the Galaxy 
formed by the accretion of protogalactic fragments that resembled our 
dwarf satellite galaxies {\em as they were $\gtrsim$~10~Gyr ago}, then 
the RR Lyrae pulsation properties in the Galactic halo and in the dwarf 
galaxies should be basically indistinguishable. The main goal of the present 
paper is to check whether this is the case or not.

\section{RR Lyrae Stars in Galactic Globular Clusters and Nearby Dwarf Galaxies} 
\label{sec:rrl}

Galactic globular clusters provide a well-known tracer of the properties of 
the Galactic halo. In the present section, we compare the properties of the 
RR Lyrae stars in Galactic globular clusters with those of RR Lyrae stars in 
globular clusters and the general field of the Milky Way dwarf 
satellites. Specifically, we compare the average periods of the ab-type 
(fundamental-mode) RR Lyrae stars in the different populations. 
The empirical data, along with extensive references, can be found in  
\citet{mc06}. 

In Figure~\ref{fig:OO} we compare the RRab period distribution of Galactic 
globular clusters and nearby extragalactic systems (the LMC, the SMC, the 
dSph satellites of the Milky Way, and their associated globular clusters). 
Clearly, the Milky Way globulars present the well-known 
{\em Oosterhoff dichotomy} \citep{o39,o44}, or the lack of systems with 
average RRab periods in the range between 0.58~d and 0.62~d. This is valid 
both for the ``old halo'' and ``young halo'' subsystems of globular clusters,
in the \citet{mvdb05} nomenclature. Note that 
the satellite distribution peaks where the Galactic distribution reaches 
a minimum. In other words, the Oosterhoff dichotomy is not present among the 
Milky Way satellite galaxies.   

We can quantify the above statements by carrying out statistical tests. 
The KMM test \citep*{kaea94} shows that the distribution of 
$\langle P_{\rm ab}\rangle$ for the Galactic globular clusters is better 
described by a bimodal rather than a unimodal distribution, with 
99.99\% confidence (or higher, depending on whether the metal-rich 
clusters NGC~6388 and NGC~6441, marked ``OoIII'' in Fig.~\ref{fig:OO}, as 
well as $\omega$~Centauri, are included or not). More specifically, the 
fit assigns 59\% of the Galactic globulars into the OoI mode, with a 
$\langle P_{\rm ab}\rangle = 0.563$~d, and 41\% of the clusters into the 
OoII mode, with $\langle P_{\rm ab}\rangle = 0.662$~d. The estimated 
common covariance is only $8.8\times 10^{-4}$. This amounts to quantitative 
proof that the Oosterhoff dichotomy is indeed present among Galactic 
globular clusters. 
Figure~\ref{fig:OO} clearly shows, in contrast, that the satellite 
systems do not primarily belong to either of these two groups: in fact, a 
fit with two Gaussians provides one mode, containing 86\% of the objects, 
that is centered right in the middle of the ``Oosterhoff gap'' zone, with 
$\langle P_{\rm ab}\rangle = 0.592$~d. These conclusions remain basically 
unchanged whether we introduce the dwarf galaxy populations (i.e., their 
field stars) or not. This demonstrates that the Oosterhoff 
gap is {\em not} present among the satellite populations.

\begin{figure*}[ht]
  \includegraphics[angle=0,scale=0.55]{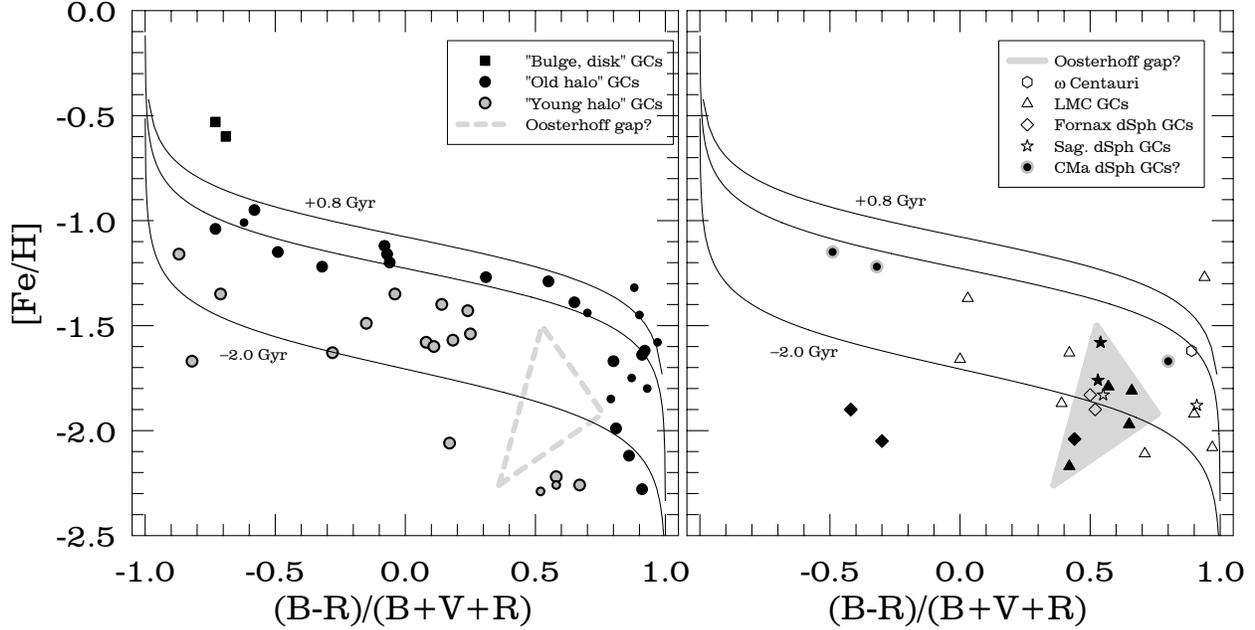} 
  \caption{({\em Left panel}) 
  Position of the RR Lyrae-bearing Galactic globular clusters 
  with a defined Oosterhoff type in the metallicity--``HB type'' plane. 
  The symbols are the same as in Figure~\ref{fig:OO}. 
  ({\em Right panel}) To the previous plot 
  the position of the RR Lyrae-bearing 
  globular clusters which have been associated with 
  dwarf satellite galaxies of the Milky Way are added. 
  Filled symbols for the extragalactic 
  systems indicate an {\em Oosterhoff-intermediate} status. Note the
  concentration of Oosterhoff-intermediate clusters in the triangular 
  region marked ``Oosterhoff gap?'' 
\label{fig:HBR}
}  
\end{figure*}

\section{Horizontal Branch Morphology: Galactic vs. Nearby Extragalactic 
  Globular Clusters} 
\label{sec:hbm}

RR Lyrae stars occupy the evolutionary phase known as the {\em horizontal 
branch} (HB) phase \citep[see][ for a review]{mc06}. As such, it is 
legitimate to ask: is there a systematic difference in HB morphology (i.e., 
in the relative proportions between red, blue, and variable HB stars) between 
the RR Lyrae-bearing Galactic globular cluster system, on the one hand, and 
the RR Lyrae-bearing nearby extragalactic globular cluster system, on the 
other, that may help explain their different Oosterhoff behaviors?  

An answer to this question is provided in Figure~\ref{fig:HBR}, which shows 
the Lee-Zinn HB type parameter $\LZ = (B\!-\!R)/(B\!+\!V\!+R)$ (where $B$, 
$R$, and $V$ are the numbers of blue, red, and variable HB stars, 
respectively) plotted as a function of the metallicity 
\citep[data from][]{mc06}. In the {\em left-hand panel}, only the Galactic 
globular clusters are shown, overplotted on isochrones from \citet{cfp93}. 
The {\em right-hand plot}, in turn, shows the position of the globular 
clusters associated with the dwarf satellite galaxies of the Milky Way. 
In the latter panel, Oosterhoff-intermediate clusters are shown as filled 
symbols. Note that the nearby extragalactic globular clusters tend to clump 
around a triangular region in this plane, which is basically empty in the 
case of the Galactic system \citep{mc06}. 

The fact that the nearby extragalactic globular clusters appear to clump 
around a region of the $\LZ - {\rm [Fe/H]}$ plane where basically 
no Galactic globular clusters can be found again suggests that the two 
systems may be profoundly different. We test this 
hypothesis by performing a 2-dimensional, 2-sample Kolmogorov-Smirnov 
test, as described in \S14.7 in \citet{wpea92}. Comparing the sample of 
globular clusters associated with the Milky Way dwarf satellite galaxies 
with the bona-fide Galactic globular clusters gives a probability of 
99.4\% that the distributions have been drawn from a different parent 
population. Removing the metal-rich globular clusters NGC~6388 and 
NGC~6441 from the Galactic sample still gives a probability of 98.9\% 
that the two distributions are inconsistent. 

It has at times been 
suggested that the globular clusters in the Milky Way's dwarf satellite 
galaxies may resemble the so-called ``young halo'' population
\citep[e.g.,][]{mvdb05}. This is 
not borne out by these statistical tests, which give a probability of 
99.5\% that the so-called ``young halo'' and dwarf galaxy-related 
globular clusters have been drawn from different parent populations.

\section{Cosmological Implications}
\label{sec:cos}

What constraints do the above results pose on the formation history of the 
Galaxy? 

In terms of the $\Lambda$CDM paradigm, the preceding discussion strongly suggests 
that the ``protogalactic fragments'' that may have given rise to the Galactic 
halo must have had little to do with even the {\em very early} Sagittarius, 
Fornax, or LMC dwarf galaxies. In other words, if the Galaxy had formed from 
accretion of galaxies resembling the aforementioned ones, even its {\em oldest 
stellar populations}, as traced by the RR Lyrae stars, would have looked 
significantly different from what is currently observed. 

By indicating that, even at the very beginning, these dwarf galaxies must have 
looked fundamentally different from any protogalactic fragments that may have 
helped build the Milky Way halo, RR Lyrae stars clearly allow us to push even  
further the previous constraints on the role played by dwarf galaxies in the 
formation of the Galactic halo.  

\begin{figure}[t]
  \includegraphics[angle=0,scale=0.35]{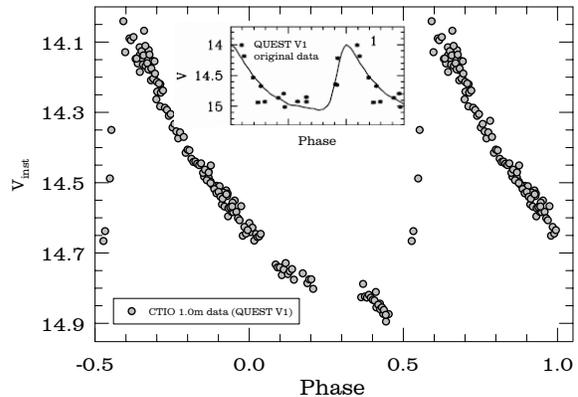} 
  \caption{
  Light curve for QUEST variable 1, as determined from data collected at 
  the CTIO 1.0m telescope (Smith et al. 2006, in preparation), compared 
  against the original QUEST light curve for the same star (inset). Note 
  the ill-defined light curve minimum and maximum in the QUEST curve, which 
  renders the amplitude---and hence the star's Oosterhoff type---rather 
  uncertain. 
\label{fig:QUEST}
}  
\end{figure}

As far as the general halo field, the situation is a bit more complicated. 
While \citet*{nsea91} have strongly argued, from a careful analysis of the 
light curves of halo RR Lyrae stars, that the Oosterhoff dichotomy {\em is} 
indeed present in the halo field 
\citep[see also Fig.~1, right panel, in][]{mc04}, other authors have 
recently questioned these conclusions on the basis of datasets drawn from  
sky surveys. As an example, the QUEST survey \citep{kv06} has led to the 
discovery of many previously uncatalogued RR Lyrae stars. Based on these 
data, \citet{vz03} have challenged the \citeauthor{nsea91} results, 
claiming that 
the Oosterhoff dichotomy is {\em not} seen among their sample stars. While 
it remains unclear what the final solution to this baffling discrepancy  
will be, we note that many of the Oosterhoff-intermediate stars in the 
QUEST database \citep{avea04} have very incomplete light curves, which 
make their classification into an Oosterhoff group highly uncertain. 
In fact, in the case of individual field stars and dwarf galaxies, a 
reliable Oosterhoff classification 
must be based also on a period-amplitude diagram, which requires the 
rejection of stars showing the Blazhko effect \citep{mc04}. As an example, 
Figure~\ref{fig:QUEST} shows a comparison between RR Lyrae variable 1 in the 
QUEST catalog and the new (instrumental) lightcurve in $V$ that we have 
recently obtained for the same star. In this particular case, it is clear 
that the QUEST and our amplitudes differ substantially, in the right sense 
to move the star away from the Oosterhoff-intermediate region of the 
period-amplitude diagram and into the OoII region. More comprehensive surveys 
of halo RR Lyrae stars will be required before we can confidently rule out 
(or confirm) the general validity of the \citeauthor{nsea91} results. 
When we do, we will be able to place additional constraints on the extent 
to which such dwarf satellite galaxies as Sculptor, Draco, Carina, and 
Ursa Minor, all of which lack globular clusters, may have taken part in 
the very early formation history of the Galactic halo.

\acknowledgements

I thank T. D. Kinman, R. Salinas, H. A. Smith, and K. A. Vivas for useful 
discussions. This work has been supported by Fondecyt grant \#1030954.

\end{document}